\documentclass[aps,pra,twocolumn,superscriptaddress,10pt,nofootinbib,longbibliography,floatfix]{revtex4-2}
\usepackage{ORI_Group_Style}
\usepackage{symbols}
\usepackage[english]{babel}

\usepackage{relsize}
\usepackage{xcolor}
\usepackage{scalerel}
\usepackage{tikz}
\usetikzlibrary{svg.path}
\usepackage{soul}
\definecolor{orcidlogocol}{HTML}{A6CE39}
\tikzset{
  orcidlogo/.pic={
    \fill[orcidlogocol] svg{M256,128c0,70.7-57.3,128-128,128C57.3,256,0,198.7,0,128C0,57.3,57.3,0,128,0C198.7,0,256,57.3,256,128z};
    \fill[white] svg{M86.3,186.2H70.9V79.1h15.4v48.4V186.2z}
                 svg{M108.9,79.1h41.6c39.6,0,57,28.3,57,53.6c0,27.5-21.5,53.6-56.8,53.6h-41.8V79.1z M124.3,172.4h24.5c34.9,0,42.9-26.5,42.9-39.7c0-21.5-13.7-39.7-43.7-39.7h-23.7V172.4z}
                 svg{M88.7,56.8c0,5.5-4.5,10.1-10.1,10.1c-5.6,0-10.1-4.6-10.1-10.1c0-5.6,4.5-10.1,10.1-10.1C84.2,46.7,88.7,51.3,88.7,56.8z};
  }
}
\newcommand\orcid[1]{\href{https://orcid.org/#1}{\mbox{\scalerel*{
\begin{tikzpicture}[yscale=-1,transform shape]
\pic{orcidlogo};
\end{tikzpicture}
\ 
}{|}}}}

\hypersetup{colorlinks,linkcolor=blue,citecolor=blue,urlcolor=blue}

\begin{document}

\newcommand{\inns}{\affiliation{Institute for Quantum Optics and Quantum Information of the Austrian Academy of Sciences, 6020 Innsbruck, Austria}
\affiliation{Institute for Theoretical Physics, University of Innsbruck, 6020 Innsbruck, Austria}}

\title{Numerical Simulation of Large-Scale Nonlinear Open Quantum Mechanics}

\author{M.~Roda-Llordes} 
\author{D.~Candoli}
\inns
\author{P.~T.~Grochowski} 
\inns
\affiliation{Center for Theoretical Physics, Polish Academy of Sciences, Aleja Lotnik\'ow 32/46, 02-668 Warsaw, Poland}
\author{A.~Riera-Campeny} 
\author{T.~Agrenius} 
\inns
\author{J.~J.~García-Ripoll}  
\affiliation{Instituto de Física Fundamental IFF-CSIC, Calle Serrano 113b, Madrid 28006, Spain}
\author{C.~Gonzalez-Ballestero} 
\author{O.~Romero-Isart} 
\inns

\date{\today}

\begin{abstract}

We introduce a numerical method to simulate nonlinear open quantum dynamics of a particle in situations where its state undergoes significant expansion in phase space while generating small quantum features at the phase-space Planck scale. Our approach involves simulating the Wigner function in a time-dependent frame that leverages information from the classical trajectory to efficiently represent the quantum state in phase space. To demonstrate the capabilities of our method, we examine the open quantum dynamics of a particle evolving in a one-dimensional weak quartic potential after initially being ground-state cooled in a tight harmonic potential. This numerical approach is particularly relevant to ongoing efforts to design, optimize, and understand experiments targeting the preparation of macroscopic quantum superposition states of massive particles through nonlinear quantum dynamics.

\end{abstract}

\maketitle

\section{Introduction}

The field of levitodynamics~\cite{Gonzalez-Ballestero2021_Levitodynamics}, which focuses on levitation and control of microobjects in vacuum, allows us to study the center-of-mass motional dynamics of a particle in a highly isolated environment. Since the mechanical potential in which the particle moves can be controlled both dynamically~\cite{Ciampini2021_Experimental,Rakhubovsky2021_Stroboscopic,Neumeier2022_Fast} and statically~\cite{Pino2018_Onchip,Roda-Llordes2023_Macroscopic}, levitated particles offer a unique platform to study nonlinear conservative mechanics. 
Furthermore, the center-of-mass thermal energy can be removed, either via active or passive feedback, to the ultimate limit where only quantum fluctuations are present~\cite{Delic2020_Cooling,Magrini2021_Realtime,Tebbenjohanns2021_Quantum,Kamba2022_Optical,Ranfagni2022_Twodimensional,Piotrowski2022_Simultaneous,Kamba2023_Nanoscale}. Center-of-mass ground-state cooling and the control of the mechanical potential open up the possibility to study nonlinear quantum mechanics with a microsolid containing billions of atoms~\cite{Moore2022_Hierarchy,Roda-Llordes2023_Macroscopic}. In order to design, optimize, and understand experimentally feasible protocols involving nonlinear quantum mechanics, it is crucial to have a reliable numerical tool that allows us to efficiently simulate the dynamics while accounting for sources of noise and decoherence. In this paper we provide such a tool in the particularly relevant and challenging scenario of multiscale dynamics induced by center-of-mass cooled massive particles evolving in wide nonharmonic potentials.

More specifically, the center-of-mass motion of cooled microparticles exhibits minute fluctuations (i.e., zero-point motion), smaller than the size of a single atom. Experimentally feasible nonharmonic potentials are wider than zero-point motion length scales, that is the distance between classical turning points is orders of magnitude larger than the zero-point length scale. Hence, the dynamics triggered in those nonharmonic potential will generate large phase-space expansions. This expansive dynamics will eventually activate the nonharmonicities in the potential, such as at turning points, which in the case of coherent dynamics can create phase-space structures at or even below the Planck scale~\cite{Zurek2001_SubPlanck}. This multiscale phase-space dynamics of the particle's center-of-mass state will be studied through the time evolution of the corresponding Wigner function. The use of the Wigner function is advantageous as it enables us to incorporate sources of noise and decoherence (i.e., open dynamics) while also clearly identifying quantum features (e.g., through negative values in the Wigner function).
To effectively describe the scenario of interest, which involves large phase-space expansions and small phase-space features and is thus different from previous studies~\cite{Cabrera2015_Efficient}, an efficient numerical representation of this specific dynamics is necessary. We propose using a time-dependent phase-space grid where the grid points move according to the classical trajectory dictated by the nonharmonic potential. This procedure places the grid points where they are most relevant, thereby improving computational efficiency. We call this numerical tool {\em Q-Xpanse}, and it has proven invaluable in the design, optimization, and understanding of a recent proposal for generating macroscopic quantum superpositions of a nanoparticle through the nonlinear quantum mechanics induced in a wide double-well potential~\cite{Roda-Llordes2023_Macroscopic}.

This paper is structured as follows: In Section~\ref{sec:flowing_frame}, we present the theoretical framework for our method, including the time-dependent change of variables leading to the time-dependent phase-space grid. In Section~\ref{sec:numerical_implementation} and in a dedicated Appendix section, we detail our numerical implementation using finite differences and classical trajectory propagation. We then examine the dynamics in weak quartic potentials as an example of large expansions with Planck-scale quantum features in Section~\ref{sec:example}. Finally, we conclude with our final remarks and outlook in Section~\ref{sec:conclusion}.

\section{Wigner function dynamics in the Liouville frame} \label{sec:flowing_frame}

We consider a particle with mass $\mass$ evolving in a one-dimensional potential $\potential(\x)$ in the presence of noise.
We describe the state of the particle through its Wigner function $W(\x,\p,t)$.
The equation of motion for the Wigner function is given by
\begin{equation} \label{eq:Wigner_FPE}
    \partialfrac[W(\x,\p,t)]{t}  = \pare{\classicL + \quantumL + \noiseL} W(\x,\p,t).
\end{equation}
The first term generates conservative (i.e., Liouville) classical dynamics and is given by
\begin{equation}
    \classicL = - \frac{\p}{\mass}\partialfrac{\x} + \partialfrac[\potential(\x)]{\x} \partialfrac{\p}.
\end{equation}
The second term generates genuine quantum dynamics and is given by
\begin{equation}
    \quantumL = \sum_{n=1}^\infty \frac{(-1)^{n}}{(2n+1)!} \frac{\hbar^{2n}}{4^n}
    \frac{\partial^{2n+1} \potential(\x)}{\partial \x^{2n+1}} \frac{\partial^{2n+1}}{\partial \p^{2n+1}}.
\end{equation}
 Note that $\quantumL$ is zero for quadratic potentials (i.e., potentials with only linear and harmonic terms). The third term models the presence of noise and generates dissipative dynamics. For levitated nanoparticles, it is convenient to consider~\cite{Romero-Isart2011_Quantum} 
 \begin{equation} \label{eq:noiseL}
    \noiseL = \friction \pare{ 1 + \p \partialfrac{\p} } +  \frac{\hbar^2 \dnoiserate}{2\xzpf^2} \partialfrac[^2]{\p^2},
\end{equation}
where
\begin{equation}
\dnoiserate = \frac{\friction k_\text{B} T}{\hbar\freq} + \dnoiserate_1,
\end{equation}
 $k_\text{B}$ is the Boltzmann constant, and $\xzpf=\spares{\hbar/(2\mass\freq)}^{1/2}$ is a convenient length unit associated to the zero-point motion fluctuations of the quantum ground state of a harmonic potential with frequency $\freq$.
This source of noise models a linear coupling to a thermal bath~\cite{Breuer2010_Theory} of temperature $T$, with damping rate $\gamma$, and the presence of a stochastic white-force term with displacement noise rate given by $\Gamma_1$. 

The results of this paper are based on using the Wigner function in a time-dependent frame that we call the {\em Liouville frame}, which is defined as
\begin{equation} \label{eq:Wflow_definition}
    \flowW(\x,\p,t) \equiv e^{-\classicL t} W(\x,\p,t).
\end{equation}
Since $\classicL$ is the generator of classical dynamics, one can use the Liouville theorem to write 
\begin{align} \label{eq:Wflow_as_W}
    \flowW(\x,\p,t) = W(\xc(\x,\p,t),\pc(\x,\p,t),t),
\end{align}
where $\xc(\x,\p,t)$ and $\pc(\x,\p,t)$ are the solutions to the classical equations of motion for point particles moving in the potential $\potential(\x)$ in the absence of noise with initial position and momentum given by $\x$ and $\p$ respectively. Namely, they are solutions of
\begin{equation} \label{eq:classical_ODE}
\begin{split}
   \frac{ \partial \xc(\x,\p,t)}{\partial t} &= \frac{\pc(\x,\p,t)}{\mass},
    \\
    \frac{\partial \pc(\x,\p,t)}{\partial t} &= -\left. \frac{\partial \potential(x)}{\partial x} \right |_{x=\xc(\x,\p,t)},
\end{split}
\end{equation}
with $\xc(\x,\p,0) = \x$ and $\pc(\x,\p,0) = \p$. In the Liouville frame, the Wigner function evolves as
\begin{equation} \label{eq:Wigner_FPE_flowing}
    \partialfrac[\flowW(\x,\p,t)]{t}  =  e^{-\classicL t}  \pare{\quantumL + \noiseL}  e^{\classicL t} \, \flowW(\x,\p,t).
\end{equation}
The Wigner function in the Liouville frame evolves only due to the presence of quantum effects and/or noise, that is $\partial \flowW(\x,\p,t)/\partial t =0$ if $\quantumL = \noiseL=0$. The original Wigner function $W(\x,\p,t)$ can be obtained from the Wigner function in the Liouville frame $\flowW(\x,\p,t)$ by
\begin{equation} \label{eq:W_as_Wflow}
    \begin{split}
    W(\x,\p,t) &=  e^{\classicL t} \flowW(\x,\p,t) \\ &= \flowW(\xc(\x,\p,-t),\pc(\x,\p,-t),t),
    \end{split}
\end{equation}
that is, by using backward propagation in time of the classical trajectories.

\begin{figure*}[t]
    \centering
    \includegraphics[width=\linewidth]{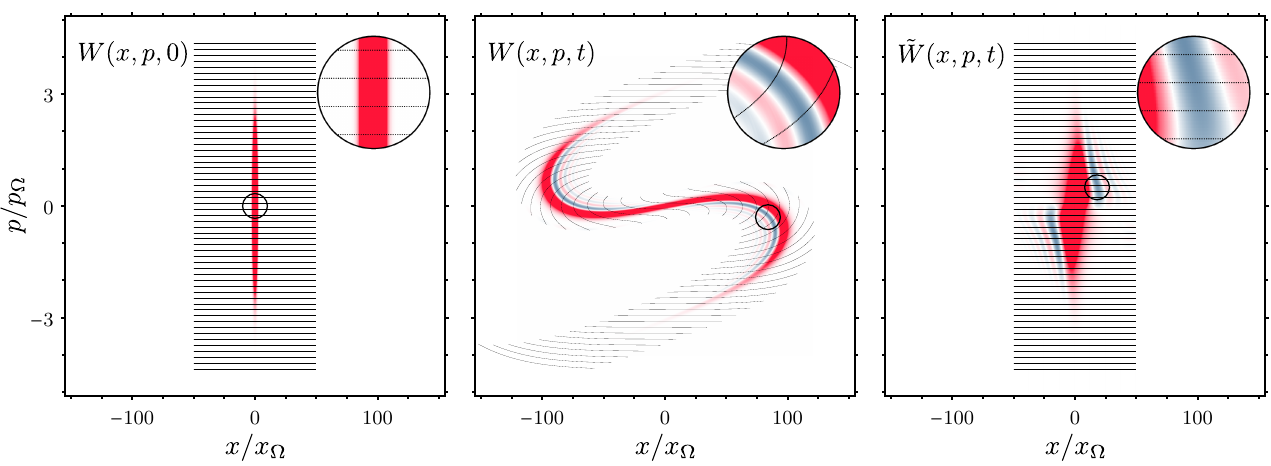}    
    \caption{Evolution in time of the Wigner function for a particle initially prepared in the ground state of the harmonic potential $\potentialHarmonic(x)$ and evolving until $\freq t=150$  in the quartic potential $\potentialQuartic(x)$ (see \eqnref{eq:quarticpotential}) with $\eta=10^2$ in the presence of decoherence with $\dnoiserate = 10^{-5} \freq$. Left panel shows the initial state $W(\x,\p, 0) = \flowW(\x,\p, 0)$, middle and right panel show the evolved state in the original frame $W(\x,\p, 150/\freq)$ and in the Liouville frame  $\flowW(\x,\p, 150/\freq)$, respectively. 
    The black points, which appear as lines due to their high density, depict a regular grid in the Liouville frame which we used to simulate the dynamics.
    The grid has $255\times 56$ points with $h_\x/\xzpf \approx 0.39$ and $h_\p / \pzpf \approx 0.16$.}
    \label{fig:1}
\end{figure*}

In the following section, we will show that numerically solving \eqnref{eq:Wigner_FPE_flowing} on a fixed regular phase-space grid is highly efficient in situations involving large expansions because it corresponds to solving \eqnref{eq:Wigner_FPE} on a time-dependent, irregular phase-space grid that places grid points where they are most crucial. This key idea is illustrated in \figref{fig:1} for the example  of a particle evolving in a pure quartic potential which we will further discuss in \secref{sec:example}.

\section{Numerical Simulation in the Liouville Frame} \label{sec:numerical_implementation}

In this section we explain how to numerically solve the time evolution of the Wigner function in the Liouville frame, namely how to solve \eqnref{eq:Wigner_FPE_flowing}. The first step is to explicitly calculate the terms in $e^{-\classicL t}  \pare{\quantumL + \noiseL}  e^{\classicL t}$. This allows us to obtain the explicit form of the partial derivative equation (PDE). As shown in \appref{app:num-details}, one obtains that \eqnref{eq:Wigner_FPE_flowing} reads 
\begin{equation} \label{eq:Wigner_FPE_flowing_FULL}
    \partialfrac[\flowW (\x,\p,t)]{t} = \sum_{n,m=0}^{n+m \leq N_U} \auxFunc_{nm}(\x,\p,t) \frac{\partial^{n+m} \flowW (\x,\p,t)}{ \partial \x^n \partial \p^m}.
\end{equation}
Here $N_U$ is the smallest odd number such that $\partial^{n} \potential(\x)/\partial \x^n = 0$ for $n \ge N_U+2$, which in turn determines that \eqnref{eq:Wigner_FPE_flowing_FULL} is a PDE of order $N_U$.
The time-dependent scalar functions $\auxFunc_{nm}(\x,\p,t)$ depend on the physical parameters of the problem (i.e., $\mass$, $\potential(\x)$, $\gamma$, $T$, $\Gamma_1$), both explicitly and implicitly through the classical trajectories $\xc(\x,\p,t)$ and $\pc(\x,\p,t)$ and their up to $N_U$ order derivatives with respect to their initial condition $\p$.
Their derivation and explicit expressions for an up to quartic potential ($N_U=3$) are given in \appref{app:num-details}.

The second step is to convert the PDE in \eqnref{eq:Wigner_FPE_flowing_FULL} into a system of linear equations using the method of finite differences.
In the Liouville frame we use a uniform rectangular grid in $\x$ and $\p$ with separation between consecutive grid points given by $h_\x>0$ and $h_\p>0$ along each direction respectively. The grid points are given by $(\x_i,\p_j) = (x_0,p_0) + (i h_\x,j h_\p)$, for $i=0,1,\ldots,N_\x-1$ and $j=0,1,\ldots,N_\p-1$. Here $(\x_0,\p_0)$ is the bottom left point of the grid which contains $N=N_\x\times N_\p$ points. The $N$ values of the Wigner function in the Liouville frame $\flowW (\x,\p,t)$ evaluated at the grid points are collected by the $N$-dimensional vector $\vectflowW(t)$ whose components, indexed by $k=0,1,\ldots,N-1$, are given by $\flowW_{k=iN_\p +j}(t) = \flowW(\x_i,\p_j,t)$. Using a finite difference method (see \appref{app:num-details} for further details), one obtains a system of linear equations for this vector given by
\begin{equation} \label{eq:linear-fd-ode}
    \partialfrac[\vectflowW (t)]{t} = \matrixD(t) \vectflowW(t),
\end{equation}
where $\matrixD(t)$ is a square $N \times N$ matrix. The equation \eqnref{eq:linear-fd-ode} can then be solved using 
\begin{equation} \label{eq:exp-propagate}
    \vectflowW(t+\Delta t) = \exp \spare{\matrixD(t) \Delta t} \vectflowW(t),
\end{equation}
which is valid for a sufficiently small $\Delta t$ (see \appref{app:num-details}).

This numerical method relies on developing a numerically efficient way of computing $\matrixD(t)$, which requires evaluating $\auxFunc(\x,\p,t)$ at the grid points.
In turn, this requires the values of $\xc(\x,\p,t)$ and $\pc(\x,\p,t)$ as well as the derivatives of $\xc(\x,\p,-t)$ and $\pc(\x,\p,-t)$ with respect to $\p$ at every grid point $(\x_i,\p_j)$ and for all instances of time considered in the finite differences approach.
Since an analytical formula for the classical trajectories and their derivatives with respect to initial conditions are in general not available for nonharmonic potentials, they need to be efficiently evaluated numerically. We do this by solving \eqnref{eq:classical_ODE} and similar differential equations that can be derived for the derivatives of the classical trajectories with respect to initial conditions using a symplectic method, which ensures stability over long integration times~\cite{Sivak2014_Time}. Another important tool we use to improve the efficiency of the method is to relate the derivatives of the forward-propagated trajectories with respect to the initial conditions with those of the backward-propagated trajectories by making use of the properties of the associated Jacobian matrices. We provide all the details of this numerical method in \appref{app:num-details}, a method that we have coded using \texttt{C++}, \texttt{Cython} and \texttt{Python}.

Formally, solving \eqnref{eq:Wigner_FPE_flowing_FULL} in the fixed grid given by the $N$ phase-space points $(\x_i,\p_j)$ defined above is equivalent to solving \eqnref{eq:Wigner_FPE} in a time-dependent grid given by the $N$ points $(\x_i(t),\p_j(t)) \equiv(\xc(\x_i,\p_j,t),\pc(\x_i,\p_j,t))$, see \figref{fig:1}. In this time-dependent grid, a feature which will be important for our later discussion is the maximum phase-space grid density, namely the minimal distance between two phase-space points. In order to quantify this phase-space density let us introduce the following dimensionless Jacobian matrix for a given phase-space point, namely
\begin{equation}
    \dimJ(\x,\p,t) \equiv \begin{pmatrix}
        \dpartialfrac[\xc (\x,\p,t)]{\x} & \dpartialfrac[\xc (\x,\p,t)]{\p}\dfrac{\pzpf}{\xzpf} \\
        \dpartialfrac[\pc (\x,\p,t)]{\x}\dfrac{\xzpf}{\pzpf} & \dpartialfrac[\pc (\x,\p,t)]{\p}
    \end{pmatrix}
\end{equation}
with $\pzpf \equiv \hbar/(2\xzpf)$.
We then define $\lambda^+_{i,j}(t)$ and $\lambda^-_{i,j}(t)$ as the largest and smallest singular values of the $2\times2$ matrix $\dimJ(\x_i,\p_j,t)$. One can then define $\gridSep(t) \equiv \min_{i,j} \lambda^-_{i,j}(t)$, namely the minimum singular value over all the grid. In this way the dimensionless parameter $1/\gridSep(t)$ quantifies the maximum density of the phase-space time-dependent grid. To see this explicitly consider a point in phase space, written without dimensions as $\mathbf{r}=(\x/\xzpf,\p/\pzpf)$ and a point $\mathbf{r'} = \mathbf{r} + \epsilon (\cos \theta, \sin \theta)$ in its close vicinity (i.e., $ |\epsilon| \ll 1$).
After the evolution governed by the classical trajectory, the separation between these two points can be expressed, in linear order in $\epsilon$, as
\begin{equation} \label{eq:argument_singular_values}
    \frac{|\flowmap(\mathbf{r'},t) - \flowmap(\mathbf{r},t)|}{\epsilon} \approx |\dimJ(\x,\p,t) (\cos \theta, \sin \theta)^T| 
    \end{equation}
According to the singular value decomposition of $\dimJ(\x,\p,t)$, the smallest singular value of $\dimJ(\x,\p,t)$ minimizes the distance \eqnref{eq:argument_singular_values} over all possible directions, namely $\theta$.

\section{Example: Quartic potential} \label{sec:example}

Let us now apply the numerical method presented in this paper to a particular example: quantum mechanics in a purely quartic potential~\cite{Weiss2019_Quantum}. This will allow us to show the applicability of the numerical method and illustrate that solving $\flowW(\x,\p,t)$ in a constant and regular grid is equivalent to solving $W(\x,\p,t)$ in a {\em smart} time-dependent irregular phase-space grid, see \figref{fig:1}. 

We consider a particle of mass $\mass$, whose state at $t=0$, namely $W(\x,\p,0) = \flowW(\x,\p,0)$, is given by the ground state of the harmonic potential $\potentialHarmonic(\x) = \mass\freq^2 \x^2/2$, see left panel of \figref{fig:1}. The position and momentum standard deviation of the initial state are given by $\sqrt{\avg{\xop^2}  } =\xzpf = \sqrt{\hbar/(2\mass\freq)}$ and $\sqrt{\avg{\pop^2}  }= \pzpf = \hbar/(2\xzpf)$ respectively. At $t>0$, the particle evolves in a purely quartic potential $\potential (\x) = \potentialQuartic(\x) $,  which we parametrize as
\begin{equation} \label{eq:quarticpotential}
    \potentialQuartic(\x)= \frac{1}{ \eta^4} \, \frac{\hbar\freq}{4} \pare{\frac{\x}{\xzpf}}^4.
\end{equation}
We consider the case of friction-less noise (e.g., dynamics in ultra-high vacuum), namely $\friction =0$ but $\dnoiserate/\freq>0$ in \eqnref{eq:noiseL}. The dimensionless parameter $\eta$ characterizes the strength of the quartic potential. Since the initial kinetic energy of the state is $\hbar \Omega/4$, the turning point $\xturning$ according to classical mechanics, defined as $\hbar \Omega/4 = \potentialQuartic(\xturning)$, is given by $\xturning/\xzpf= \eta $. Large phase-space expansions, namely states with spatial delocalization orders of magnitude larger than $\xzpf$~\cite{Weiss2021_Large}, will be thus generated for $\eta \gg 1$. 

\begin{figure}
    \centering
    \includegraphics[width=\linewidth]{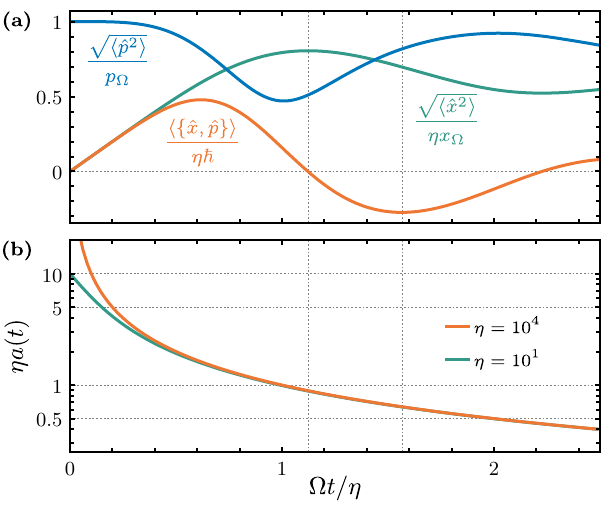}
    \caption{(a) Second order moments of a particle initially prepared in the ground state of $\potentialHarmonic(x)$ and evolving in $\potentialQuartic(x)$ with $\eta=10^3$ and and $\dnoiserate=2 \times 10^{-8}\freq$. The results for $\eta=10,10^2$ and $10^4$ in these normalized units is indistinguishable in the scale of the plot.
    The vertical lines indicate the instant where the $\sqrt{\avg{\xop^2} (t) }/(\eta \xzpf)$ is maximum and the instant where $\avg{\{\xop,\pop\}}(t)/(\eta \hbar)$ reaches its most negative value respectively. 
    (b) Grid density parameter $\gridSep(t)$ as a function of time for the quartic potential $\potentialQuartic(x)$ for 2 different values of $\eta$.
    }
    \label{fig:2}
\end{figure}

Let us first analyze the evolution of the first and second phase-space moments. Due to the alignment of the quartic potential with the initial state, the first moments remain constant and equal to zero, namely $\avg{\xop}(t)/\xzpf = \avg{\pop}(t)/\pzpf =0$. The dynamics of the second moments is shown in \figref{fig:2}(a), where we plot   $\sqrt{\avg{\xop^2} (t) }/(\eta \xzpf)$,  $\sqrt{\avg{\pop^2} (t) }/( \pzpf)$, and $\avg{\{\xop,\pop\}}(t)/(\eta \hbar)$. Using the $\eta$-scaled dimensionless timescale $t \freq/\eta$, this plot is for $\eta \gg 1$ conveniently independent of $\eta$. The plot shows that the state experiences free dynamics during an initial time scale given by $0 <  t \freq/\eta \lesssim 0.4 $, where  $\sqrt{\avg{\xop^2} (t) }/(\eta \xzpf) \approx \avg{\{\xop,\pop\}}(t)/(\eta \hbar)$ grows linearly in time and $\sqrt{\avg{\pop^2} (t) }/( \pzpf)$ remains constant and equal to one. After this initial time interval, the state starts to experience the quartic potential. In particular, at $t \freq/\eta \approx 1.12$, when $\avg{\{\xop,\pop\}}(t)/(\eta \hbar)$ is equal to zero, the state reaches a maximum value of $\sqrt{\avg{\xop^2} (t) }/(\eta \xzpf)$ of the order of 1, that is, the state is spatially delocalized to a large length scale given by $\eta \xzpf$~\cite{Weiss2021_Large}. This expansive dynamics that generates a squeezed state is conveniently accompanied with an increase of the phase-space grid density. This can be shown in \figref{fig:2}(b), where we plot the $\eta$-scaled grid distance $\eta \gridSep(t)$ as a function of time, showing that the phase-space density grows as a function of time and is scaled with $\eta$.

\begin{figure}
    \centering
    \includegraphics[width=\linewidth]{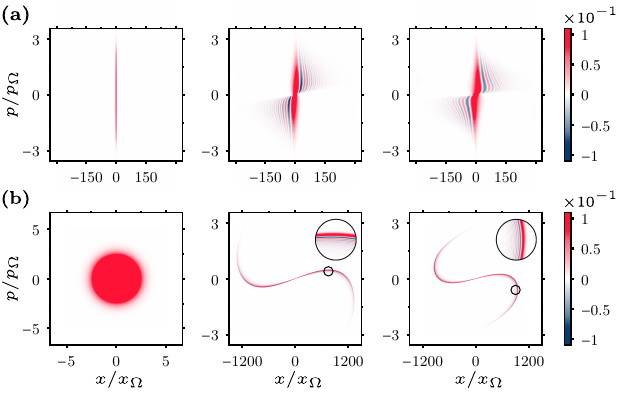}
    \caption{Phase space representation of the state of a particle initially prepared in the ground state of $\potentialHarmonic(x)$ and evolving in $\potentialQuartic(x)$ with $\eta=10^3$ and $\dnoiserate=2 \times 10^{-8}\freq$. 
    Results at different instances of time, namely the initial state, the time where the variance in position is maximum and the time when the covariance is minimum, see \figref{fig:2}.
    (a) $\flowW(\x,\p,t)$. In this case, we used a grid with $2048\times 256$ points with $h_\x/\xzpf \approx 0.24$ and $h_\p / \pzpf \approx 0.06$.
    (b) $W(\x,\p,t)$.}
    \label{fig:3}
\end{figure}

\begin{figure}
    \centering
    \includegraphics[width=\linewidth]{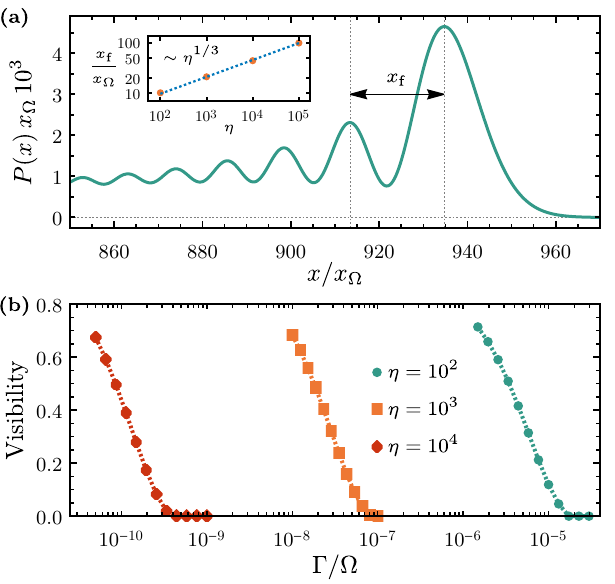}
    \caption{(a) Probability distribution in position $P(\x)$ at time $t \freq /\eta =1.56$ for a particle initially prepared in the ground state of $\potentialHarmonic(x)$ and evolving in $\potentialQuartic(x)$ with $\eta=10^3$ and $\dnoiserate/\freq=2 \times 10^{-8}$.
    This time corresponds to the moment where the covariance is minimum, see \figref{fig:2}.
    The inset shows how the separation between the first two peaks $\x_\text{f}$ scales as a function of $\eta$.
    (b) Visibility of the second largest maximum of $P(\x)$ at the time specified above, as a function of displacement noise rate $\Gamma$ and for different values of $\eta$.}
    \label{fig:4}
\end{figure}

Let us now study the evolution of the Wigner function. In \figref{fig:3} we show $\flowW(\x,\p,t)$ (panel a) and $W(\x,\p,t)$ (panel b) for $\eta = 10^3$ and $\Gamma/\freq = 2 \times 10^{-8}$ at three instances of time: (i) $t \freq =0$,  (ii) $\freq t/\eta \approx 1.12$ when $\sqrt{\avg{\xop^2} (t) }/(\eta \xzpf)$ is the largest and the state generates an interference pattern in the momentum probability distribution, and (iii) $\freq t/\eta \approx 1.56$ when $\avg{\{\xop,\pop\}}(t)/(2 \eta \hbar)$ reaches its most negative value and the state exhibits an interference pattern in the position probability distribution, see \figref{fig:4}(a). In \figref{fig:2}(a) the instances of time (ii) and (iii) are indicated with a vertical dashed line. We emphasize that $W(\x,\p,t)$ (panel b) is obtained by simply using \eqnref{eq:W_as_Wflow} after having numerically obtained $\flowW(\x,\p,t)$ (panel a) with the method presented in this paper. Comparing the $x$ axes of panels (a) and (b) of \figref{fig:3}, one can see how $W(\x,\p,t)$ expands significantly more than $\flowW(\x,\p,t)$. As shown in \figref{fig:1}, the regular grid points used to represent $\flowW(\x,\p,t)$ in the Liouville frame are efficiently distributed in the original frame to properly describe $W(\x,\p,t)$. 
The results in \figref{fig:3}(a) are obtained in a fixed grid of a phase-space length scale given by $\sqrt{\pares{h_\x/\xzpf}^2+\pares{h_\p/\pzpf}^2} \approx 0.25$. The length scale in the time-dependent grid, namely in \figref{fig:3}(b) is reduced by a factor of $\gridSep(t)$, which as one can see in \figref{fig:2}(b), reaches values below $10^{-3}$, way below the phase-space Planck scale~\cite{Zurek2001_SubPlanck}.
This means that to match the accuracy level of our method, using a regular grid in the original frame would need about $10^3$ times more grid points.

Finally, let us discuss the impact of noise by illustrating how it affects the visibility of the interference pattern in position at the time $t \freq /\eta \approx 1.56$. In \figref{fig:4}(a) we plot the probability distribution $P(x) \equiv \intall dp \, W(\x,\p,t) $ at this particular instance of time for $\eta=10^3$ and $\dnoiserate/\freq=2 \times 10^{-8}$. We define $x_f$ as the distance between the largest interference peak and its neighboring peak. For the parameters in \figref{fig:4}, we obtain $x_f/\xzpf \approx 21.2$. As shown in the inset of \figref{fig:4}(a), the scaling of this distance with $\eta$ is given by $x_f/\xzpf \approx 2.11 \eta^{1/3}$. The visibility of this interference pattern, defined as $(P_\text{max}-P_\text{min})/(P_\text{max}+P_\text{min})$ where $P_\text{max}$ and $P_\text{min}$ are the value of $P(x)$ at the largest maximum and its neighboring minimum respectively, is a decreasing a function of $\Gamma/\freq$ as we show in \figref{fig:4}(b). As expected~\cite{Weiss2021_Large,Roda-Llordes2023_Macroscopic}, the impact of $\Gamma$ in the visibility scales roughly as $\eta^2$.

The study of quantum dynamics in a nonharmonic potential in the presence of noise, which we have performed using the numerical method presented in this paper, is relevant for current efforts to prepare largely delocalized macroscopic quantum states of large masses~\cite{Romero-Isart2011_Large,Yin2013_Large,Scala2013_MatterWave,Bateman2014_Nearfield,Hoang2016_Electron,Gieseler2020_SingleSpin,Marshman2022_Constructing,Romero-Isart2017_Coherent,Neumeier2022_Fast,Roda-Llordes2023_Macroscopic}. We remark that, feasibility-wise, purely quartic potentials are not ideal since the time scale needed to generate the interference pattern shown in \eqnref{fig:4}, that is $t \freq /\eta \approx 1.56$, is for $\eta \gg 1$ much larger than the average collision time with a single gas molecule at ultra-high vacuum~\cite{Romero-Isart2011_Quantum}. This is one of the main reasons motivating our recent proposal~\cite{Roda-Llordes2023_Macroscopic}, which is also analyzed with the numerical method presented in this paper, where we use a double-well potential such that the inverted harmonic term exponentially speeds up the dynamics~\cite{Romero-Isart2017_Coherent, Pino2018_Onchip}.

\section{Conclusions} \label{sec:conclusion}

In this paper we have presented a numerical method that can simulate nonlinear open quantum dynamics, even for potentials in which the quantum state expands several orders of magnitude in phase space while exhibiting relevant features at very small sub-Planck scales~\cite{Zurek2001_SubPlanck}.
This regime is of particular interest for designing, optimizing, and understanding protocols that generate macroscopic quantum states by letting a massive particle evolve in a nonharmonic potential~\cite{Moore2019_Estimation,Rakhubovsky2021_Stroboscopic, Roda-Llordes2023_Macroscopic}.
We have demonstrated the power of this method using the dynamics of an initially highly-localized state in a quartic potential. 
We have shown how in this potential the state position variance grows by several orders of magnitude, and yet its Wigner function exhibits negative features on a scale below the initial zero-point fluctuations. Properly describing such small scales using a regular grid in the original frame would require an impracticable amount of points, a challenge that we overcome by the introduction of the Liouville frame.

Our numerical method should be applicable to a broad class of interesting quantum mechanical problems. While any potential $\potential(\x)$ can be considered, the number of derivatives considered in \eqnref{eq:Wigner_FPE_flowing} must be finite to allow for a numerical evaluation. Introducing a cutoff to the order of the potential in $\potential(\x)$ should yield accurate results.
Other types of noise and decoherence beyond the ones considered in this paper (e.g. stochastic force-gradient) can also be incorporated. While we have considered both time-independent potentials and decoherence rates, the numerical method is inherently time dependent, see \eqnref{eq:exp-propagate}, which means that time dependence could be introduced with the corresponding modifications. An advantageous feature of studying quantum mechanics with the Wigner function is that the classical limit can be easily taken, namely taking $\hbar=0$ such that  $\quantumL=0$ in \eqnref{eq:Wigner_FPE}. In this classical limit, the Wigner function in the Liouville frame is only driven by dissipative dynamics. 
Finally, while we have focused on a one-dimensional problem, the method could be generalized to higher spatial dimensions.

In conclusion, the numerical method presented in this manuscript relies on a crucial element: the description of Wigner function dynamics in the Liouville frame~\eqnref{eq:Wigner_FPE_flowing}. We emphasize that this frame proves to be highly valuable not only in practical terms but also from a conceptual standpoint, as it clearly unveils the impact of quantum physics in the mechanical motion of a particle. 

We would like to thank Christoph Dellago, Lukas Einkemmer, Daniele Giannandrea, Max Innerbichler, Talitha Weiss and the Q-Xtreme synergy group for helpful discussions.
This research has been supported by the European Research Council (ERC) under the grant agreement No. [951234] (Q-Xtreme ERC-2020-SyG) and by the European Union’s Horizon 2020 research and innovation programme under grant agreement No. [863132] (IQLev).
PTG was partially supported by the Foundation for Polish Science (FNP).

\bibliography{bibliography.bib}

\appendix
\onecolumngrid 

\section{Details on the Numerical Method} \label{app:num-details}

In this appendix we detail all the steps we use to solve \eqnref{eq:Wigner_FPE_flowing} numerically.

\subsection{Explicit expression for the PDE}

The first step is to obtain an explicit expression for $e^{-\classicL t} \pare{\quantumL + \noiseL} e^{\classicL t}$.
In order to find how any operator $\op$ transforms under $e^{-\classicL t} \op \, e^{\classicL t}$, one can apply the transformed operator to an arbitrary function $\func(\x,\p)$ and identify which operator produces the same result.
It is useful to recall that by virtue of the Liouville theorem we know how $e^{\pm\classicL t}$ acts on an arbitrary function $\func(\x,\p)$, namely
\begin{equation} \label{eq:arbitrary_liouville}
    e^{\pm\classicL t} f(\x,\p) = f(\xc(\x,\p,\mp t), \pc(\x,\p,\mp t)).
\end{equation}

In our case, the operators that appear in $\quantumL + \noiseL$ are $\x$, $\p$ and derivatives with respect to $\p$.
For $\x$ and $\p$, making use of \eqnref{eq:arbitrary_liouville} one finds that these operators transform according to
\begin{equation} \label{eq:lab_from_flow}
    e^{-\classicL t} \x \, e^{\classicL t} = \xc(\x,\p,t) \quad\text{and}\quad
    e^{-\classicL t} \p \, e^{\classicL t} = \pc(\x,\p,t).
\end{equation}
Similarly, one finds that the derivative with respect to $\p$ transforms according to the chain rule as
\begin{equation} \label{eq:chain1}
    e^{-\classicL t} \partialfrac{\p} \, e^{\classicL t} 
    = \xcInDer{1} \partialfrac{\x} + \pcInDer{1} \partialfrac{\p}.
\end{equation}
where we introduce the following shorthand notation
\begin{equation} \label{eq:inverse-deriv}
    \xcInDer{n} = \left. \partialfrac[^{n}\xc(\x,\p,-t)]{\p^n} \right|_{\substack{\x=\xc(\x,\p,t) \\ \p = \pc(\x,\p,t)}}
    \quad\text{and}\quad
    \pcInDer{n} = \left. \partialfrac[^{n}\pc(\x,\p,-t)]{\p^n} \right|_{\substack{\x=\xc(\x,\p,t) \\ \p = \pc(\x,\p,t)}}.
\end{equation}
Note that these are scalar functions of $\x$, $\p$ and $t$, and they are the derivatives with respect to initial conditions of the classical trajectories starting from the point $(\xc(\x,\p,t),\pc(\x,\p,t))$ propagated backwards in time for a time $t$. Explicitly, for $n=1$ they correspond to the following limit
\begin{equation} \label{eq:inverse_derivative_limit}
    \xcInDer{1} = \lim_{\varepsilon\rightarrow 0} \frac{\xc(\xc(\x,\p,t),\pc(\x,\p,t)+\varepsilon,-t) - \xc(\xc(\x,\p,t),\pc(\x,\p,t),-t)}{\varepsilon} = \lim_{\varepsilon\rightarrow 0} \frac{\xc(\xc(\x,\p,t),\pc(\x,\p,t)+\varepsilon,-t) - \x}{\varepsilon}.
\end{equation}
For higher order derivatives one finds expressions corresponding to multiple applications of the chain rule.
Namely, for the second order derivative with respect to $\p$ one has
\begin{equation} \label{eq:chain2}
\begin{split}
    e^{-\classicL t} \partialfrac[^2]{\p^2} \, e^{\classicL t} = 
    \xcInDer{2} \partialfrac{\x} + 
    \pcInDer{2} \partialfrac{\p} +
    \pare{\xcInDer{1}}^2 \partialfrac[^2]{\x^2} + 
    \pare{\pcInDer{1}}^2 \partialfrac[^2]{\p^2}+
    2 \xcInDer{1}\pcInDer{1}\partialfrac[^2]{\x\partial\p},
\end{split}
\end{equation}
whereas for the third order derivative one has
\begin{equation} \label{eq:chain3}
\begin{split}
    e^{-\classicL t} \partialfrac[^3]{\p^3} \, e^{\classicL t} =& \xcInDer{3} \partialfrac{\x} +
    \pcInDer{3} \partialfrac{\p} + 3 \spare{ \xcInDer{1} \xcInDer{2} \partialfrac[^2]{\x^2} + 
    \pcInDer{1} \pcInDer{2} \partialfrac[^2]{\p^2} + 
    \pare{ \xcInDer{2} \pcInDer{1} + \xcInDer{1} \pcInDer{2} } \partialfrac[^2]{\x\partial\p} } + \\
    & \pare{\xcInDer{1}}^3 \partialfrac[^3]{\x^3} + \pare{\pcInDer{1}}^3 \partialfrac[^3]{\p^3} +
    3 \spare{ \pare{\xcInDer{1}}^2\pcInDer{1} \partialfrac[^3]{\x^2\partial\p} + \xcInDer{1}\pare{\pcInDer{1}}^2 \partialfrac[^3]{\x\partial\p^2} }.
\end{split}
\end{equation}
We only consider potentials $\potential(\x)$ for which fifth and higher order derivatives vanish, and therefore only derivatives up to third order will appear in \eqnref{eq:Wigner_FPE_flowing}.
For potentials where higher order derivatives are relevant, one could extend our approach to include them.
Substituting Eqs.~\eqref{eq:lab_from_flow}, \eqref{eq:chain1}, \eqref{eq:chain2} and \eqref{eq:chain3} into \eqnref{eq:Wigner_FPE_flowing} yields the explicit equation that we need to solve numerically.
It has the following form,
\begin{equation} \label{eq:Wigner_FPE_flowing_FULL_appendix}
    \partialfrac[\flowW (\x,\p,t)]{t} = \sum_{n,m=0}^{n+m \leq 3} \auxFunc_{nm}(\x,\p,t) 
    \partialfrac[^{n+m} \flowW (\x,\p,t)]{\x^n \partial \p^m},
\end{equation}
where the explicit expressions for the coefficients are given by
\begin{align}
    \auxFunc_{00}(\x,\p,t) &= \friction  \label{eq:gStart} \\
    \auxFunc_{10}(\x,\p,t) &= \friction\, \pc \xcInDer{1} 
    + \frac{\hbar^2\dnoiserate}{2\xzpf^2} \xcInDer{2} 
    - \frac{\hbar^2}{12} \potential^{(3)}(\xc) \xcInDer{3} \\
    \auxFunc_{01}(\x,\p,t) &= \friction\, \pc(\x,\p,t) \pcInDer{1} 
    + \frac{\hbar^2\dnoiserate}{2\xzpf^2} \pcInDer{2} 
    - \frac{\hbar^2}{12} \potential^{(3)}(\xc) \pcInDer{3} \\
    \auxFunc_{20}(\x,\p,t) &= \frac{\hbar^2\dnoiserate}{2\xzpf^2} \pare{\xcInDer{1}}^2
    - \frac{\hbar^2}{4} \potential^{(3)}(\xc) \xcInDer{1}\xcInDer{2} \\
    \auxFunc_{02}(\x,\p,t) &= \frac{\hbar^2\dnoiserate}{2\xzpf^2} \pare{\pcInDer{1}}^2
    - \frac{\hbar^2}{4} \potential^{(3)}(\xc) \pcInDer{1}\pcInDer{2} \\
    \auxFunc_{11}(\x,\p,t) &= \frac{\hbar^2\dnoiserate}{2\xzpf^2} \xcInDer{1}\pcInDer{1}
    - \frac{\hbar^2}{4} \potential^{(3)}(\xc) \pare{ \xcInDer{1}\pcInDer{2} + \pcInDer{1}\xcInDer{2} } \\
    \auxFunc_{30}(\x,\p,t) &= - \frac{\hbar^2}{12} \potential^{(3)}(\xc) \pare{\xcInDer{1}}^3 \\
    \auxFunc_{03}(\x,\p,t) &= - \frac{\hbar^2}{12} \potential^{(3)}(\xc) \pare{\pcInDer{1}}^3 \\
    \auxFunc_{21}(\x,\p,t) &= - \frac{\hbar^2}{4} \potential^{(3)}(\xc) \pare{\xcInDer{1}}^2 \pcInDer{1} \\
    \auxFunc_{12}(\x,\p,t) &= - \frac{\hbar^2}{4} \potential^{(3)}(\xc) \pare{\pcInDer{1}}^2 \xcInDer{1}. \label{eq:gEnd}
\end{align}
In order to simplify notation, here and hereafter we use $\potential^{(i)}(\x)$ for the $i-$th derivative of $\potential$ evaluated at $\x$. 
Also, note that we use $\xc$ and $\pc$ as a shorthand for $\xc(\x,\p,t)$ and $\pc(\x,\p,t)$ to simplify the expressions, but they still depend on $\x$, $\p$ and $t$.

\subsection{Discretization of the PDE}

Now that we have an explicit expression for the equation we need to solve, we need to discretize it to allow for numerical simulation.
In order to do so we describe $\flowW$ in a regular grid which contains $N=N_\x \times N_\p$ points which we denote by $(\x_i,\p_i)$.
Then, we denote the values of $\flowW$ in each of these grid points by $\flowW_{i,j}=\flowW(\x_i,\p_j)$.
Next, we express the derivatives with respect to $\x$ and $\p$ in \eqnref{eq:Wigner_FPE_flowing_FULL_appendix} in terms of finite difference schemes.
In particular, we use a second-order centered finite difference scheme, which we list below for the first, second and third order derivatives.
First, for the first order derivatives they read
\begin{align} \label{eq:FDStart}
    \partialfrac[\flowW_{i,j}]{\x} &= \frac{\flowW_{i+1,j}-\flowW_{i-1,j}}{2h_\x}, \\
    \partialfrac[\flowW_{i,j}]{\p} &= \frac{\flowW_{i,j+1}-\flowW_{i,j-1}}{2h_\p}.
\end{align}
Next, for the second order derivatives one has
\begin{align}
    \partialfrac[^2\flowW_{i,j}]{\x^2} &= \frac{\flowW_{i+1,j}+\flowW_{i-1,j}-2\flowW_{i,j}}{h_\x^2}, \\
    \partialfrac[^2\flowW_{i,j}]{\p^2} &= \frac{\flowW_{i,j+1}+\flowW_{i,j-1}-2\flowW_{i,j}}{h_\p^2}, \\
    \partialfrac[^2\flowW_{i,j}]{\x \partial\p} &= \frac{\flowW_{i+1,j+1}+\flowW_{i-1,j-1} -\flowW_{i-1,j+1}-\flowW_{i+1,j-1}}{4h_\x h_\p}.
\end{align}
Finally, the expressions for the third order derivatives are given by
\begin{align}
    \partialfrac[^3\flowW_{i,j}]{\x^3} &=
        \frac{\flowW_{i+2,j}-2\flowW_{i+1,j}+2\flowW_{i-1,j}-\flowW_{i-2,j}}{2h_\x^3}, \\
    \partialfrac[^3\flowW_{i,j}]{\p^3} &=
        \frac{\flowW_{i,j+2}-2\flowW_{i,j+1}+2\flowW_{i,j-1}-\flowW_{i,j-2}}{2h_\p^3}, \\
    \partialfrac[^3\flowW_{i,j}]{\x^2 \partial \p} &= 
        \frac{ \flowW_{i+1,j+1} + \flowW_{i-1,j+1} - 2\flowW_{i,j+1} +2\flowW_{i,j-1} - \flowW_{i+1,j-1} - \flowW_{i-1,j-1}}{2h_\x^2h_\p}, \\
    \partialfrac[^3\flowW_{i,j}]{\x \partial \p^2} &= 
        \frac{\flowW_{i+1,j+1} + \flowW_{i+1,j-1} - 2\flowW_{i+1,j} + 2\flowW_{i-1,j} - \flowW_{i-1,j+1} - \flowW_{i-1,j-1} }{2h_\x h_\p^2}. 
    \label{eq:FDEnd}
\end{align}
After substituting all the derivatives in \eqnref{eq:Wigner_FPE_flowing_FULL_appendix} by their finite difference versions [see Eqs.~\eqref{eq:FDStart}--\eqref{eq:FDEnd}], the right hand side of the equation is given by a linear combination of $\flowW_{i,j}$ with different indices $i,j$. Explicitly, one has
\begin{equation} \label{eq:Wigner_FPE_flowing_discrete}
    \partialfrac[\flowW_{i,j}(t)]{t} = \sum_{\alpha} \mathcal{D}_{(i,j),(\alpha)} \flowW_{\alpha}(t)
\end{equation}
where $\alpha$ runs over the following 13 indices $(i,j)$,$(i\pm 1,j)$,$(i,j\pm 1)$,$(i\pm 1,j\pm 1)$,$(i\pm 1,j\mp 1)$,$(i\pm 2,j)$ and $(i,j\pm 2)$.
Collecting the values of $\flowW_{i,j}$ in the $N$-dimensional vector $\vectflowW(t)$ with components $\flowW_{k=iN_\p +j}(t) = \flowW_{i,j}(t)$ indexed by $k=0,1,\ldots,N-1$ allows us to write \eqnref{eq:Wigner_FPE_flowing_discrete} as
\begin{equation}
     \partialfrac[\vectflowW (t)]{t} = \matrixD(t) \vectflowW(t),
\end{equation}
where $\matrixD(t)$ is a $N\times N$ matrix. 
From this equation one can derive an expression to propagate the solution in time given by
\begin{equation}
    \vectflowW (t + \Delta t) = \exp\spare{\int_t^{t+\Delta t} \matrixD(t') dt' } \vectflowW (t)
    \approx \exp\spare{\matrixD(t) \Delta t} \vectflowW (t)
\end{equation}
where the approximation assumes that $\Delta t$ is small enough such that $\matrixD(t')$ varies slowly enough between $t$ and $t+ \Delta t$.

The entries of the $\matrixD(t)$ matrix can be found by inspection after replacing the derivatives in \eqnref{eq:Wigner_FPE_flowing_FULL_appendix} by their finite difference versions [see Eqs.~\eqref{eq:FDStart}--\eqref{eq:FDEnd}]. For instance the matrix entry corresponding to the index $(i,j),(i+1,j)$ reads
\begin{equation} \label{eq:example_D}
    \mathcal{D}_{(i,j),(i+1,j)}(t) = \frac{\auxFunc_{10}(\x_i,\p_j,t)}{2h_\x} 
    + \frac{\auxFunc_{20}(\x_i,\p_j,t)}{h^2_\x} 
    - \frac{\auxFunc_{30}(\x_i,\p_j,t)}{h^3_\x}
    - \frac{\auxFunc_{21}(\x_i,\p_j,t)}{h_\x h^2_\p}.
\end{equation}
Notice that each row of $\matrixD(t)$ will only have 13 entries different from zero, which means that $\matrixD(t)$ will be sparse.
This is due to the fact that finite differences only relate points with up to second order neighbours.
One could have chosen higher-order finite differences, in which case there would me more nonzero entries in each row of $\matrixD(t)$.
However, we found that increasing the finite differences from second to fourth order didn't yield any significant improvement in the accuracy of our solution.
Finally, note that in order to fully define $\matrixD(t)$ one needs to specify the boundary conditions.
We use periodic boundary conditions since they provide a more stable simulation than zero-value boundary conditions.
In particular, we identify the right and top edges of the grid with the left and bottom edges respectively.
Explicitly, we identify $i=N_\x$ with $i=0$, and $j=N_\p$ with $j=0$.

\subsection{Efficient computation of the $\mathcal{D}$ matrix}

As one can see in \eqnref{eq:example_D}, obtaining the numerical value for the different entries of the $\matrixD(t)$ matrix requires evaluating all $\auxFunc_{mn}(\x_i,\p_j,t)$ [see Eqs.~\eqref{eq:gStart}--\eqref{eq:gEnd}] in each point of the grid. 
In turn, this requires the values of $\xc(\x_i,\p_j,t)$ and $\pc(\x_i,\p_j,t)$ as well as the derivatives $\xcInDer{n}$ and $\pcInDer{n}$ [see \eqnref{eq:inverse-deriv}] up to $n=3$ at every grid point $(\x_i,\p_j)$ and for all instances of time considered in the finite differences approach.
Since an analytical formula for the classical trajectories is generally not available for nonharmonic potentials we evaluate them numerically.

We obtain $\xc(\x_i,\p_j,t)$ and $\pc(\x_i,\p_j,t)$ by propagating in time the classical equations of motion \eqnref{eq:classical_ODE} with each grid point $(\x_i,\p_j)$ as initial condition.
To ensure stability over long integration times we use a symplectic method~\cite{Sivak2014_Time}. 
In particular, we use the 4-th order method described in~\cite{Yoshida1990_Construction}.
To obtain the derivatives of the inverse mapping $\xcInDer{n}$ and $\pcInDer{n}$, we use an approach consisting of two steps.
First, we compute the derivatives of the direct mapping as solutions to differential equations, which allows us to benefit from the properties of the symplectic method used above.
Second, we use these values to compute $\xcInDer{n}$ and $\pcInDer{n}$ through the relation between the direct and inverse mapping. 
Using these steps is more efficient than a direct numerical evaluation of these derivatives in terms of limits such as the one shown in \eqnref{eq:inverse_derivative_limit}.
In the following we describe these two steps in detail.

By taking derivatives with respect to $\x$ and $\p$ in \eqnref{eq:classical_ODE} one can obtain the equation of motion for the derivatives we need.
Note that we use $\xc$ and $\pc$ as a shorthand for $\xc(\x,\p,t)$ and $\pc(\x,\p,t)$ respectively.
Specifically, taking the derivative with respect to $\x$ on \eqnref{eq:classical_ODE} yields the differential equations for $\partial_{\x} \xc(\x,\p,t)$ and $\partial_{\x} \pc(\x,\p,t)$
\begin{equation} \label{eq:classical_flowx_ODE}
\begin{split}
    \partial_t \partialfrac[\xc]{\x} &= \frac{1}{\mass} \partialfrac[\pc]{\x},
    \\
    \partial_t \partialfrac[\pc]{\x} &= -\potential^{(2)}(\xc) \partialfrac[\xc]{\x}.
\end{split}
\end{equation}
The initial conditions are given by $\partial_{\x} \xc(\x,\p,0) = 1$ and $\partial_{\x} \pc(\x,\p,0)= 0$.
They stem from the fact that, at time $t=0$, $\xc(\x,\p,0)=\x$ and $\xc(\x,\p,0)=\p$.
Similarly, taking the derivative with respect to $\p$ yields a similar equation for $\partial_{\p} \xc(\x,\p,t)$ and $\partial_{\p} \pc(\x,\p,t)$,
\begin{equation} \label{eq:classical_flowp_ODE}
\begin{split}
    \partial_t \partialfrac[\xc]{\p} &= \frac{1}{\mass} \partialfrac[\pc]{\p},
    \\
    \partial_t \partialfrac[\pc]{\p} &= -\potential^{(2)}(\xc) \partialfrac[\xc]{\p},
\end{split}
\end{equation}
with initial conditions $\partial_{\p} \xc(\x,\p,0) = 0$ and $\partial_{\p} \pc(\x,\p,0)= 1$.
By taking more derivatives, one can obtain equations for the higher order derivatives.
The second order derivatives with respect to initial conditions fulfill
\begin{equation} \label{eq:classical_flowx2_ODE}
\begin{split}
    \partial_t \partialfrac[^2\xc]{\x^2} &= \frac{1}{\mass} \partialfrac[^2\pc]{\x^2},
    \\
    \partial_t \partialfrac[^2\pc]{\x^2} &= 
    -\potential^{(3)}(\xc) \pare{\partialfrac[\xc]{\x}}^2
    -\potential^{(2)}(\xc) \partialfrac[^2\xc]{\x^2},
\end{split}
\end{equation}
\begin{equation} \label{eq:classical_flowp2_ODE}
\begin{split}
    \partial_t \partialfrac[^2\xc]{\p^2} &= \frac{1}{\mass} \partialfrac[^2\pc]{\p^2},
    \\
    \partial_t \partialfrac[^2\pc]{\p^2} &= 
    -\potential^{(3)}(\xc) \pare{\partialfrac[\xc]{\p}}^2
    -\potential^{(2)}(\xc) \partialfrac[^2\xc]{\p^2},
\end{split}
\end{equation}
\begin{equation} \label{eq:classical_flowxp_ODE}
\begin{split}
    \partial_t \partialfrac[^2\xc]{\x\partial \p} &= \frac{1}{\mass} \partialfrac[^2\pc]{\x\partial \p},
    \\
    \partial_t \partialfrac[^2\pc]{\x\partial \p} &= 
    -\potential^{(3)}(\xc) \partialfrac[\xc]{\x}\partialfrac[\xc]{\p}
    -\potential^{(2)}(\xc) \partialfrac[^2\xc]{\x\partial \p},
\end{split}
\end{equation}
with all the initial conditions being zero.
The third order derivatives fulfill
\begin{equation} \label{eq:classical_flowx3_ODE}
\begin{split}
    \partial_t \partialfrac[^3\xc]{\x^3} &= \frac{1}{\mass} \partialfrac[^3\pc]{\x^3},
    \\
    \partial_t \partialfrac[^3\pc]{\x^3} &= 
    -\potential^{(4)}(\xc) \pare{\partialfrac[\xc]{\x}}^3
    -3\,\potential^{(3)}(\xc) \partialfrac[\xc]{\x} \partialfrac[^2\xc]{\x^2}
    -\potential^{(2)}(\xc) \partialfrac[^3\xc]{\x^3},
\end{split}
\end{equation}
\begin{equation} \label{eq:classical_flowp3_ODE}
\begin{split}
    \partial_t \partialfrac[^3\xc]{\p^3} &= \frac{1}{\mass} \partialfrac[^3\pc]{\p^3},
    \\
    \partial_t \partialfrac[^3\pc]{\p^3} &= 
    -\potential^{(4)}(\xc) \pare{\partialfrac[\xc]{\p}}^3
    -3\,\potential^{(3)}(\xc) \partialfrac[\xc]{\p} \partialfrac[^2\xc]{\p^2}
    -\potential^{(2)}(\xc) \partialfrac[^3\xc]{\p^3},
\end{split}
\end{equation}
\begin{equation} \label{eq:classical_flowxxp_ODE}
\begin{split}
    \partial_t \partialfrac[^3\xc]{\x^2\partial \p} &= \frac{1}{\mass} \partialfrac[^3\pc]{\x^2\partial \p},
    \\
    \partial_t \partialfrac[^3\pc]{\x^2\partial \p} &= 
    -\potential^{(4)}(\xc) \pare{\partialfrac[\xc]{\x}}^2\partialfrac[\xc]{\p}
    -\potential^{(3)}(\xc) \partialfrac[^2\xc]{\x^2}\partialfrac[\xc]{\p}
    -2\,\potential^{(3)}(\xc) \partialfrac[\xc]{\x}\partialfrac[^2\xc]{\x\partial \p}
    -\potential^{(2)}(\xc) \partialfrac[^3\xc]{\x^2\partial \p},
\end{split}
\end{equation}
\begin{equation} \label{eq:classical_flowxpp_ODE}
\begin{split}
    \partial_t \partialfrac[^3\xc]{\x\partial \p^2} &= \frac{1}{\mass} \partialfrac[^3\pc]{\x\partial \p^2},
    \\
    \partial_t \partialfrac[^3\pc]{\x\partial \p^2} &= 
    -\potential^{(4)}(\xc) \partialfrac[\xc]{\x} \pare{\partialfrac[\xc]{\p}}^2
    -\potential^{(3)}(\xc) \partialfrac[\xc]{\x}\partialfrac[^2\xc]{\p^2}
    -2\,\potential^{(3)}(\xc) \partialfrac[\xc]{\p}\partialfrac[^2\xc]{\x\partial \p}
    -\potential^{(2)}(\xc) \partialfrac[^3\xc]{\x\partial \p^2},
\end{split}
\end{equation}
again with all the initial conditions being zero.

Note that $\xc$ and $\pc$ appear explicitly in all the equations.
Similarly, $\partial_{\x} \xc$, $\partial_{\p} \xc$, $\partial_{\x} \pc$ and $\partial_{\p} \pc$ appear in the equations for the second and third order derivatives, and the second order derivatives appear in the equations for the third order derivatives.
This means that in order to solve the equations for higher order derivatives, the values for all the lower derivatives are needed as an input.
Even more, not only the values at each time $t$ being considered are needed, but also the values at the 4 intermediate time steps in the 4-th order method \cite{Yoshida1990_Construction} that we use.
In order to be memory efficient, we do not use a separate solver for each equation, but rather use a single solver for all the equations that correctly uses all the previously computed values in the right sequence.

Finally, we need to relate these derivatives to the derivatives of the inverse map $\xcInDer{n}$ and $\pcInDer{n}$.
For the first order derivatives, the key observation is that the Jacobian matrix of the map 
\begin{equation}
    \flowJ(\x,\p,t) \equiv \begin{pmatrix}
        \dpartialfrac[\xc (\x,\p,t)]{\x} & \dpartialfrac[\xc (\x,\p,t)]{\p} \\
        \dpartialfrac[\pc (\x,\p,t)]{\x} & \dpartialfrac[\pc (\x,\p,t)]{\p},
    \end{pmatrix}.
\end{equation}
is by construction the inverse of the Jacobian matrix of the inverse map
\begin{equation} \label{eq:inv_Jacobian_matrix}
    \Ja(\x,\p,t) = 
    \begin{pmatrix}
        \partial_\x \xcInv & \xcInDer{1} \\
        \partial_\p \xcInv & \pcInDer{1}
    \end{pmatrix}.
\end{equation}
Using this fact, we compute $\flowJ(\x_i,\p_j,t)$ for each point in the grid at each time step, and then obtain $\Ja$ by inverting the matrix.
Explicitly, we use the following formula $\Ja(\x_i,\p_j,t) = \flowJ^{-1}(\x_i,\p_j,t)$.
One can show that the determinant of both $\Ja$ and $\flowJ$ is constant and equal to one, and therefore, computing this inverse is straightforward.
Similar relationships exist for higher order derivatives, which we derive below.

In order to simplify the expressions in the following, we will define the vector $\pos = (\x,\p)$ and the vector function $\flowmap(\pos,t) = (\xc(\pos,t),\pc(\pos,t))$.
Finally, we will define a new set of variables $\flowpos = (\flowx,\flowp)$ which are related to $\pos$ through the classical trajectories as
\begin{equation}
    \pos = \flowmap(\flowpos,t) \qquad\text{ or equivalently }\qquad \flowpos = \flowmap(\pos,-t).
\end{equation}
Using this notation, we can express the Jacobian matrices discussed above as
\begin{equation}
    \flowJ^i_{j} = \partialfrac[\pos_i]{\flowpos_j}
    \quad\text{and}\quad
    \Ja^i_{j} = \partialfrac[\flowpos_i]{\pos_j},
\end{equation}
and their relationship of being the inverse of each other as $\sum_k \flowJ^i_k \Ja^k_j = \delta_{ij}$.
Now, to derive a relation for the second order derivatives, we start by defining the Hessian tensor and inverse Hessian tensor respectively as
\begin{equation}
    \flowH^i_{jk} = \partialfrac[^2\pos_i]{\flowpos_j \partial \flowpos_k}
    \quad\text{and}\quad
    \He^i_{jk} = \partialfrac[^2\flowpos_i]{\pos_j \partial \pos_k}
\end{equation}
where $i,j,k$ can be either 1 or 2.
Next, we expand the following expression using the chain rule
\begin{equation} \label{eq:second-order-trick}
    0 = \partialfrac[^2 \pos_i]{\pos_j\partial \pos_k}
    = \partialfrac{\pos_k} \pare{ \sum_l \partialfrac[\pos_i]{\flowpos_l} \partialfrac[\flowpos_l]{\pos_j} }
    = \sum_l \partialfrac[\pos_i]{\flowpos_l} \partialfrac[^2\flowpos_l]{\pos_j\partial \pos_k}
    + \sum_{l,m} \partialfrac[^2\pos_i]{\flowpos_l \partial \flowpos_m} \partialfrac[\flowpos_l]{\pos_j} \partialfrac[\flowpos_m]{\pos_k}.
\end{equation}
Then, using the properties of the Jacobian matrices, we can rewrite the expression above as
\begin{equation}
    0 = \sum_l \flowJ^i_l \He^l_{j,k} + \sum_{l,m} \flowH^i_{l,m} \Ja^l_j \Ja^m_k
    \qquad\text{or}\qquad
    \He^i_{j,k} = - \sum_{n,l,m} \flowH^n_{l,m} \Ja^i_n \Ja^l_j \Ja^m_k.
\end{equation} \label{eq:formula-inv-hessian}
One can then use this expression to obtain the values of $\He$ in terms of $\flowH$ (which we compute by solving the differential equations described above) and the values of $\Ja$ that we already computed.

For the third order derivatives one can proceed in a similar fashion. One defines the tensors
\begin{equation}
    \Te^i_{jk\alpha} = \partialfrac[^3\flowpos_i]{\pos_j \partial \pos_k \partial \pos_\alpha} \quad\text{and}\quad
    \flowT^i_{jk\alpha} = \partialfrac[^3\pos_i]{\flowpos_j \partial \flowpos_k \partial \flowpos_\alpha}
\end{equation}
and takes yet another derivative with respect to $\pos_\alpha$ in \eqnref{eq:second-order-trick}.
Then, proceeding in a similar way, one finally arrives at the expression
\begin{equation}
    \Te^i_{j,k,\alpha} = -\sum_{n,l,m,\beta} \flowT^n_{l,m,\beta} \Ja^i_n \Ja^l_j \Ja^m_k \Ja^\beta_\alpha
    - \sum_{n,l,m} \flowH^{n}_{l,m} \Ja^i_n \pare{ \He^l_{j,k} \Ja^m_\alpha + \He^l_{j,\alpha} \Ja^m_k + \He^l_{k,\alpha} \Ja^m_j }.
\end{equation}

In summary, our numerical approach to solve \eqnref{eq:Wigner_FPE_flowing_FULL} consists of the following steps for each time step $\Delta t$.
First, propagate in time the classical trajectories, and its derivatives with respect to initial conditions, for each point in the grid.
Second, use these derivatives to compute the corresponding derivatives of the inverse map.
Third, use all these newly computed values to generate the matrix $\matrixD(t)$.
Finally, use \eqnref{eq:exp-propagate} to compute $\vectflowW$ at the new time step in terms of the values at the previous time step.
Repeating this procedure allows us to propagate $\vectflowW$ in time.
We implemented all these steps by developing our own simulation code in \texttt{C++}, \texttt{Cython} and \texttt{Python}.

\end{document}